# Interaction between two exposures: determining odds ratios and confidence intervals for risk estimates.

**Jesse Huang[1*], Ingrid Kockum[1*], Pernilla Stridh[1]**

1 Karolinska Neuroimmunology and Multiple Sclerosis Center, Center for Molecular Medicine, Department of Clinical Neuroscience, Karolinska University Hospital, Karolinska Institutet, Stockholm SE-171 77, Sweden

Corresponding author: Ingrid Kockum, Ingrid.kockum@ki.se

\* Equal contribution

## Abstract

In epidemiological research, it is common to investigate the interaction between risk factors for an outcome such as a disease and hence to estimate the risk associated with being exposed for either or both of two risk factors under investigation. Interactions can be estimated both on the additive and multiplicative scale using the same regression model. We here present a review for calculating interaction and estimating the risk and confidence interval of two exposures using a single regression model and the relationship between measures, particularly the standard error for the combined exposure risk group.

## Introduction

In epidemiological research, it is common to investigate the interaction between risk factors for an outcome such as a disease. This is done for a variety of reasons; it could for example be to maximize the benefit of public health interventions by identifying at-risk groups or to better understand the etiology of disease [1]. Interactions can be investigated on different scales, usually on the additive and multiplicative scale [1]. It is recommended to investigate both [2]. Usually, multiplicative interaction is estimated by fitting an interaction term between two exposures in a regression model [3]. Additive interaction is usually estimated by fitting separate risk models associated with individual and double exposures from which different measures of interaction can be estimated [4]. It is however also possible to estimate these measures of interaction from the same regression model that is used for estimating interaction on the multiplicative scale which has been recommended by VanderWeele and Knol [2]. An advantage with this approach is that both additive and multiplicative interaction can be estimated at the same time and by adjusting for covariates in the same way making the estimates comparable. We here present a review for calculating interaction and estimating the risk and confidence interval of two exposures using a single regression model and the relationship between measures, particularly the standard error for the combined exposure risk group.

# 1. The relationship between the risk and multiplicative interaction estimates for two exposures.

In a standard case-control cohort with a binary outcome Z = {0,1} and two binary exposures X, Y = {0,1}, the frequency of all combinations can be represented as in **Table 1**.

**Table 1. Frequency distribution of two dichotomous exposures.**

|  |  | Outcome (Z) | |
| --- | --- | --- | --- |
| Exposure 1 (X) | Exposure 2 (Y) | 0 (-) | 1 (+) |
| 1 (+) | 1 (+) | $d_0$ | $d_1$ |
| 1 (+) | 0 (-) | $c_0$ | $c_1$ |
| 0 (-) | 1 (+) | $b_0$ | $b_1$ |
| 0 (-) | 0 (-) | $a_0$ | $a_1$ |

The risk association between the outcome and exposures can be estimated using the logistic regression model shown below.

[**Equation 1.1**] $\quad Z \sim \beta_0 + \beta_X ( X ) + \beta_Y ( Y ) + \beta_{XY} ( X * Y )$

The odds ratio (OR) can be derived from the natural exponential of the regression coefficients. With $\beta_X$, $\beta_Y$, and $\beta_{XY}$ corresponding to the effects of the added risk due to only the first exposure (X, $OR_{10}$), added risk due to the second exposure (Y, $OR_{01}$), and the multiplicative interaction (MI) term, respectively.

[**Equation 1.2**] $\quad OR_{10} = \exp( \beta_X )$

[**Equation 1.3**] $\quad OR_{01} = \exp( \beta_Y )$

[**Equation 1.4**] $\quad MI = \exp( \beta_{XY} )$

The risk due to the combined exposure of both X and Y, $OR_{11}$, is similarly the exponential of $\beta_{X+Y}$, which is equal to the sum of the regression coefficients of the first exposure ( $\beta_X$ ), second exposure ( $\beta_Y$ ), and multiplicative term ( $\beta_{XY}$ ).

[**Equation 1.5**] $\quad \beta_{X+Y} = \beta_X + \beta_Y + \beta_{XY}$

[**Equation 1.6**] $\quad OR_{11} = \exp( \beta_{X+Y} )$

In addition, the 95% confidence intervals (CI) can be calculated using the corresponding standard error (SE) for each estimate.

[**Equation 1.7**]     $CI_{10} = \exp(\beta_X \pm 1.96 * SE_X)$

[**Equation 1.8**]     $CI_{01} = \exp(\beta_Y \pm 1.96 * SE_Y)$

[**Equation 1.9**]     $CI_{MI} = \exp(\beta_{XY} \pm 1.96 * SE_{XY})$

[**Equation 1.10**]    $CI_{11} = \exp(\beta_{X+Y} \pm 1.96 * SE_{X+Y})$

Measures of additive interaction as described by Rothman [5] includes the relative excess risk due to interaction (RERI), attributable proportion due to interaction (AP), and synergy index (SI), which can be estimated with the following [6]:

[**Equation 1.11**]    $RERI = OR_{11} - OR_{10} - OR_{01} + 1$

[**Equation 1.12**]    $AP = RERI / OR11$

[**Equation 1.13**]    $SI = (OR_{11} - 1) / [(OR_{10} - 1) + (OR_{01} - 1)]$

**Example 1. Example of an R-based logistic regression analysis (glm), Equation 1.1**

```
> summary(glm(Z~X+Y+X*Y,data=test,family="binomial"))

Call:
glm(formula = Z ~ X + Y + X * Y, family = "binomial", data = test)

Deviance Residuals:
   Min      1Q  Median      3Q     Max
-1.223  -1.217   1.133   1.138   1.195

Coefficients:
              Estimate Std. Error z value Pr(>|z|)
(Intercept)  0.0930904  0.1246494   0.747    0.455
X           -0.1345902  0.1792823  -0.751    0.453
Y            0.0007283  0.1766264   0.004    0.997
X:Y          0.1469936  0.2533262   0.580    0.562

(Dispersion parameter for binomial family taken to be 1)

    Null deviance: 1385.3  on 999  degrees of freedom
Residual deviance: 1384.4  on 996  degrees of freedom
AIC: 1392.4

Number of Fisher Scoring iterations: 3
```

{Obtained from using a simulated data (n=1000) with randomly generated distributions.}

## 2. The relationship between the effect estimate and standard error with the frequency distribution.

The effect estimate and standard error can also be determined in relation to the frequency data from **Table 1**. Odds ratios are defined by the ratio of the outcome (positive:negative) among the exposed over the unexposed. For example, $\{ OR_{11} = ( d_1 / d_0 ) / ( a_1 / a_0 ) \}$.

This means **Equation 1.6** can be written as:

[**Equation 2.1**]     $\beta_{X+Y} = \ln ( ( d_1 / d_0 ) / ( a_1 / a_0 ) )$

The same can be derived for the coefficients, $\beta_X$ and $\beta_Y$ .

[**Equation 2.2**]     $\beta_X = \ln ( ( c_1 / c_0 ) / ( a_1 / a_0 ) )$

[**Equation 2.3**]     $\beta_Y = \ln ( ( b_1 / b_0 ) / ( a_1 / a_0 ) )$

The intercept ( $\beta_0$ ) is defined by the reference group 0/0 (a) .

[**Equation 2.4**]     $\beta_0 = \ln ( a_1 / a_0 )$

Similarly, standard errors (SE) can be calculated in relation to the frequency table.

[**Equation 2.5**]     $SE_X = \sqrt{ 1/c_1 + 1/c_0 + 1/a_1 + 1/a_0 }$

[**Equation 2.6**]     $SE_Y = \sqrt{ 1/b_1 + 1/b_0 + 1/a_1 + 1/a_0 }$

The standard error of the double exposure estimate, $\beta_{X+Y}$ , ( $SE_{X+Y}$ ) is then:

[**Equation 2.7**]     $SE_{X+Y} = \sqrt{ 1/d_1 + 1/d_0 + 1/a_1 + 1/a_0 }$

In addition, the standard error of the intercept ( $SE_0$ ) is determined by,

[**Equation 2.8**]     $SE_0 = \sqrt{ 1/a_1 + 1/a_0 }$

## 3. Standard error of the combined exposure risk estimate

Although the standard errors are typically given for $\beta_X$ ($OR_{10}$) and $\beta_Y$ ($OR_{01}$), both the estimate and standard error for the combined exposure ($\beta_R$ and $SE_R$, respectively) is not typically provided. The estimate can be derived using **Equation 1.5**. However, we can derive the equation for calculating $SE_R$ using the typical estimates given (See **Figure 1**).

Starting with **Equation 2.7**, the standard error of the combined exposure risk ( $SE_{X+Y}$ ) is:

$$SE_{X+Y} = \text{sqrt} ( 1/d_1 + 1/d_0 + 1/a_1 + 1/a_0 )$$

From **Equation 2.8**, we know that $\{1/a_1 + 1/a_0 = ( SE_0 )^2\}$, therefore

[**Equation 3.1**] $\qquad SE_{X+Y} = \text{sqrt} ( 1/d_1 + 1/d_0 + ( SE_0 )^2 )$

We can derive $d_0$ and $d_1$ from **Equation 2.1** and **2.4**.

$\qquad \beta_{X+Y} = \ln [ ( d_1 / d_0 ) / ( a_1 / a_0 ) ]$      * **Equation 2.1**

$\qquad \beta_0 = \ln ( a_1 / a_0 )$      * **Equation 2.4**

$\qquad d_1 / d_0 = \exp( \beta_{X+Y} ) * \exp( \beta_0 )$

$\qquad d_1 / d_0 = \exp( \beta_{X+Y} + \beta_0 )$

[**Equation 3.2**] $\qquad d_0 = d_1 / \exp( \beta_{X+Y} + \beta_0 )$

$\qquad n_{11} = d_0 + d_1$

$\qquad n_{11} = ( d_1 / \exp( \beta_{X+Y} + \beta_0 ) ) + d_1$

$\qquad n_{11} = d_1 * ( 1 / \exp( \beta_{X+Y} + \beta_0 ) + 1 )$

[**Equation 3.3**] $\qquad d_1 = n_{11} / ( 1 / \exp( \beta_{X+Y} + \beta_0 ) + 1 )$

Replace $d_0$ and $d_1$ in **Equation 3.1** with **Equation 3.2-3**.

$$SE_{X+Y} = \text{sqrt} ( 1/d_1 + 1/( d_1 / \exp( \beta_R + \beta_0 ) ) + ( SE_0 )^2 )$$

$$SE_{X+Y} = \text{sqrt} ( ( 1 + \exp( \beta_R + \beta_0 ) ) / d_1 + ( SE_0 )^2 )$$

$$SE_{X+Y} = \sqrt{(1 + \exp(\beta_R + \beta_0))/(n_{11}/(1/\exp(\beta_R + \beta_0) + 1)) + (SE_0)^2}$$

$$SE_{X+Y} = \sqrt{(1 + \exp(\beta_R + \beta_0)) * (1/\exp(\beta_R + \beta_0) + 1)/n_{11} + (SE_0)^2}$$

[**Equation 3.4**]    $$SE_{X+Y} = \sqrt{(\exp(\beta_R + \beta_0) + 1/\exp(\beta_R + \beta_0) + 2)/n_{11} + (SE_0)^2}$$

[**Equation 3.5**]    $$SE_{X+Y} = \sqrt{(J + 1/J + 2)/n_{11} + (SE_0)^2}$$

$$\text{where } J = \exp(\beta_{X+Y} + \beta_0) = \exp(\beta_X + \beta_Y + \beta_{XY} + \beta_0)$$

$$= OR_{X+Y} * OR_0 = OR_{XY} * OR_X * OR_Y * OR_0$$

Using the measures from Example 1, we can calculate both the estimate and the standard error of the combined exposure ($SE_{X+Y}$) given that the number of samples with both exposures ($n_{11}$=245).

$$\beta_{X+Y} = \beta_X + \beta_Y + \beta_{XY}$$

$$\beta_{X+Y} = -0.1345902 + 0.0007283 + 0.1469936$$

$$\beta_{X+Y} = 0.0131317$$

$$J = \exp(\beta_X + \beta_Y + \beta_{XY} + \beta_0)$$

$$J = \exp(-0.1345902 + 0.0007283 + 0.1469936 + 0.0930904)$$

$$J = 1.112069$$

$$SE_{X+Y} = \sqrt{(J + 1/J + 2)/n_{11} + (SE_0)^2}$$

$$SE_{X+Y} = \sqrt{(1.112069 + 1/1.112069 + 2)/n_{11} + (0.1246494)^2}$$

$$SE_{X+Y} = \sqrt{(1.112069 + 1/1.112069 + 2)/245 + (0.1246494)^2}$$

$$SE_{X+Y} = 0.178634$$

We can check this value by performing the same logistic regression model using the derived variable, $T_{11}$, defined by:

$$T_{11} = \{1 \mid \text{if } (X=1 \text{ \& } Y=1); 0 \mid \text{if } (X=0 \text{ \& } Y=0)\}$$

**Example 2. Calculating the combined exposure risk estimate and standard error using R**

```
Call:
glm(formula = Z ~ T, family = "binomial", data = test)

Deviance Residuals:
   Min      1Q  Median      3Q     Max
-1.223  -1.217   1.133   1.138   1.138

Coefficients:
             Estimate Std. Error z value Pr(>|z|)
(Intercept)   0.09309    0.12465   0.747    0.455
T             0.01313    0.17863   0.074    0.941

(Dispersion parameter for binomial family taken to be 1)

    Null deviance: 696.06  on 502  degrees of freedom
Residual deviance: 696.06  on 501  degrees of freedom
  (497 observations deleted due to missingness)
AIC: 700.06

Number of Fisher Scoring iterations: 3
```

# Conclusion

In this overview, we detailed the relationship between risk measures and multiplicative interaction along with the method of assessing both the risk and confidence interval associated with each exposure group using the same regression model. The simplified process provides easier implementation and can be more resourceful in studies requiring repetitive calculations, such as when investigating gene x gene interactions which often consists of large number exposure pairs or when estimating significance of interactions using bootstrapping which requires numerous resampling.